# Spin-orbit interaction of light and diffraction of polarized beams

Aleksandr Ya. Bekshaev

Odessa I.I. Mechnikov National University
Dvorianska 2, 65082 Odessa, Ukraine
E-mail: bekshaev@onu.edu.ua

The edge diffraction of a homogeneously polarized light beam is studied theoretically based on the paraxial optics and Fresnel-Kirchhoff approximation, and the dependence of the diffracted beam pattern of the incident beam polarization is predicted. If the incident beam is circularly polarized, the trajectory of the diffracted beam centre of gravity experiences a small angular deviation from the geometrically expected direction. The deviation is parallel to the screen edge and reverses the sign with the polarization handedness; it is explicitly calculated for the case of a Gaussian incident beam with plane wavefront. This effect is a manifestation of the spin-orbit interaction of light and can be interpreted as a revelation of the internal spin energy flow immanent in circularly polarized beams. It also exposes the vortex character of the weak longitudinal field component associated with the circularly polarized incident beam.

Keywords: spin-orbit interaction, spin Hall effect of light, diffraction, spin momentum, optical vortex, longitudinal field component

## 1. Introduction

Apparently, evolution of propagating light beams is performed independently of its polarization: the spatial and polarization degrees of freedom are separated. However, a more detailed investigation reveals that the polarization state affects the propagating beam spatial pattern. The corresponding effects, generally rather fine and only observable under special conditions, are known as manifestations of the spin-orbit interaction (SOI), or spin-orbit coupling of light [1–19]. In many cases, SOI takes place in presence of inhomogeneous or anisotropic media and are mediated by the light–matter interaction [6–14]; however, even the free space transformation demonstrates distinct polarization-dependent features [15–19]. Among diverse manifestations of the SOI, the special place belongs to the spin Hall effect (SHE) of light – a polarization-induced transverse shift of the beam trajectory. Usually it is expressed by the fact that the 'center of gravity' (CG) of the transverse energy distribution in the beam slightly deviates from the geometric expectations, depending on the handedness of the beam circular polarization [6–18]. There are different versions of this phenomenon; it can be associated with the strong inhomogeneity occurring, e.g., at a plane boundary between different optical media [7–12], or may evolve gradually during the beam propagation through an inhomogeneous medium [1,6,13]. The most impressive are the SOI manifestations observable in freely propagating optical fields (in particular, upon tight focusing of a perturbed Gaussian beam with broken circular symmetry [15–17]).

Many features of the SHE show a remarkable analogy with the phenomena typical to the orbit-orbit coupling (interaction between the intrinsic and extrinsic spatial degrees of freedom) of light [20–28]. According to [20,26,28,29], the 'orbital' Hall effect of light appears as an external manifestation of the internal energy circulation existing in the beams with optical vortices (OV) and characterized by the intrinsic orbital angular momentum (OAM) of such beams. Distinct orbital analogs of the SOI are noticed in the OV beam propagation in media [20,21], in its refraction and reflection, etc. [22–27]. But especially impressive manifestations of the intrinsic energy flows were revealed in cases of the OV beam diffraction [4,30–37] where the diffracted beam evolution behind the diffraction screen spectacularly exposes the presence of the transverse energy circulation and enables to determine some of its quantitative parameters. Corresponding effects are independent of the beam polarization and are efficiently studied based on the scalar beam model taking into account only the transverse field components.

The mentioned analogy inspires the search for similar energy-flow mechanisms underlying the SHE [28,29]. Actually, the specific polarization-dependent internal energy flows do exist; it is the "spin current" associated with the spin momentum (Belinfante's momentum) of electromagnetic field [38–41]. It shows a circulatory character even in non-vortex beams with regular structure (e.g., Gaussian ones) provided that they are elliptically or circularly polarized [4,38]. The handedness and intensity of this 'spin' circulation are dictated by the polarization handedness and the beam amplitude inhomogeneity (an example of the spin flow distribution within the cross section of a circularly polarized Gaussian beam is schematically shown in figure 1). Additionally, such beams possess 'hidden' vorticity of its spatial structure: the circular polarization of the transverse field induces an OV in the small longitudinal component (LC) of the electromagnetic field [4,11,15,19] that always exist in any light beam. Handedness of this OV is also determined by the polarization handedness, and in spite of the fact that the LC is usually neglected in the paraxial context, one can reasonably expect that diffraction of this component will carry all signs of the usual OV diffraction. As a result, the total diffracted field should acquire some spin-dependent modifications of its spatial structure which, however small, reflect the incident beam polarization handedness, and the latter can be deduced from the diffraction pattern, just as the sign of OAM can be deduced from the diffraction of OV beams. In this paper, we theoretically verify this hypothesis, develop the means for its consistent description and characterization, and present its simple quantitative estimations and physical interpretations.

## 2. Diffraction model

In figure 1, the usual scheme of the edge diffraction is presented [34–36]. Our subject is the paraxial monochromatic (wavenumber $k$) light beam propagating along axis $z$, that diffracts at the rectilinear edge of the opaque screen situated in the plane $z = 0$. Although the effects we are studying in this paper are associated with the paraxiality violation, we still employ the paraxial approach as it provides good qualitative and, in many cases, satisfactory quantitative description. In this approximation the beam electromagnetic field can be represented as a superposition of $x$- and $y$-polarized components with electric and magnetic vectors [4,11,38,42]

$$\begin{Bmatrix} \mathbf{E}_x \\ \mathbf{H}_x \end{Bmatrix} = \exp(ikz)\left(\begin{Bmatrix} \mathbf{e}_x \\ \mathbf{e}_y \end{Bmatrix} u_x + \frac{i}{k}\mathbf{e}_z \begin{Bmatrix} \partial/\partial x \\ \partial/\partial y \end{Bmatrix} u_x\right),$$

$$\begin{Bmatrix} \mathbf{E}_y \\ \mathbf{H}_y \end{Bmatrix} = \exp(ikz)\left(\begin{Bmatrix} \mathbf{e}_y \\ -\mathbf{e}_x \end{Bmatrix} u_y + \frac{i}{k}\mathbf{e}_z \begin{Bmatrix} \partial/\partial y \\ -\partial/\partial x \end{Bmatrix} u_y\right) \quad (1)$$

where $u_j$ are the slowly varying complex amplitudes, $\mathbf{e}_j$ are the unit vectors of the coordinate axes ($j = x, y, z$). Note that in contrast to the previously considered scalar model of diffraction [34–36], now we deal with the vector electromagnetic field whose characteristic feature is the non-zero LCs $E_z = \exp(ikz)v_z$, $H_z = \exp(ikz)v_{Hz}$ with complex amplitudes

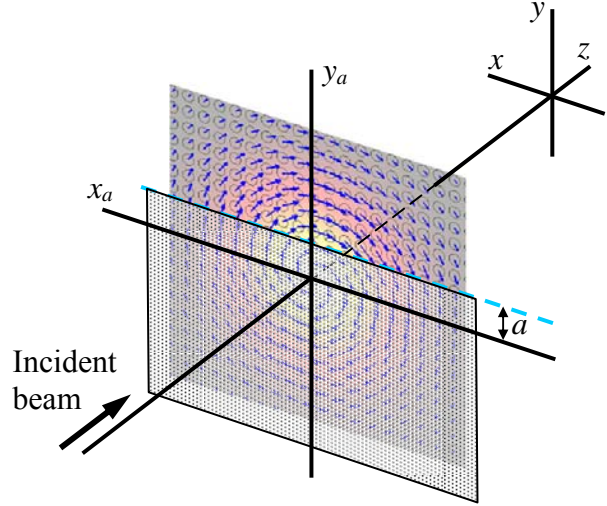

**Figure 1**. Scheme illustrating the geometrical conditions of diffraction. The opaque screen (conventionally shown as semitransparent) covers the lower part of the beam cross section ($y < a$ half-plane). The incident beam is axially symmetric and circularly polarized ($\sigma = +1$), the spin flow lines and the polarization ellipses (circles) are shown over the intensity distribution (color-coded background). The diffracted beam is observed at a distance $z$ behind the screen.

$$v_z = \frac{i}{k}\left(\frac{\partial u_x}{\partial x} + \frac{\partial u_y}{\partial y}\right), \quad v_{Hz} = \frac{i}{k}\left(-\frac{\partial u_y}{\partial x} + \frac{\partial u_x}{\partial y}\right). \quad (2)$$

Within the frame of the paraxial approximation, the longitudinal field (2) is small in respect to the transverse field,

$$v_z \sim \gamma(u_x, u_y), \quad (3)$$

where the small parameter $\gamma$ coincides with the angle of self-diffraction (beam divergence) [4,42]. Usually, the LCs (2) are discarded in paraxial optics but their existence, sometimes implicit, is of principal value and underlies some of the most fundamental properties of paraxial beams, e.g., their divergence [42,43]. In our present consideration these LC terms are also very important and lead to another fundamental property that we are going to elucidate.

In further reasoning we will deal mainly with the electric field, keeping in mind that in case of necessity the magnetic field characteristics can be easily obtained via equations (1) and (2). To make the analysis more direct, we just take the most interesting case when the incident beam is homogeneously circularly polarized, that is, for $z \leq 0$

$$u_y = i\sigma u_x \tag{4}$$

where $\sigma = \pm 1$ is the polarization handedness. Also, we suppose that the beam is axially symmetric with respect to $z$:

$$u_x(x, y, z) = u_x(r, z), \quad \frac{\partial u_x}{\partial \phi} \equiv 0 \quad (z \leq 0) \tag{5}$$

where the polar coordinates $(r, \phi)$ are introduced via the usual definitions $x = r\cos\phi$, $y = r\sin\phi$. Under conditions (4) and (5), the complex amplitudes (2) of the longitudinal field accept the form

$$v_z = \frac{i}{k}\exp(i\sigma\phi)\frac{\partial u_x}{\partial r}, \quad v_{Hz} = -i\sigma v_z \tag{6}$$

which explicitly indicates its vortex nature [4,11] mentioned in the Introduction. In this way, the analogy to the beams with OAM finds a distinct quantitative argument: at least the LC of the circularly polarized beam (1) contains the transverse energy circulation [4] that should reveal itself in the edge-diffraction processes.

The standard way of diffraction analysis is based on the Fresnel-Kirchhoff integral [44]. In the paraxial beam analysis, it is usually applied to the transverse field components $u_x$ and $u_y$ which in the Kirchhoff approximation diffract independently (the small difference in the behavior of the field components parallel and orthogonal to the screen edge that appear in the rigorous diffraction theory and depends on the screen nature [44] is neglected). For the scheme presented in figure 1 this means

$$u_{x,y}(x, y, z > 0) = \frac{k}{2\pi i z}\int_a^\infty dy_a \int_{-\infty}^\infty dx_a\, u_{x,y}(x_a, y_a, 0)\exp\left\{\frac{ik}{2z}\left[(x-x_a)^2 + (y-y_a)^2\right]\right\} \tag{7}$$

where $u_{x,y}(x_a, y_a, 0)$ is the complex amplitude of the incident beam component in the screen plane. Once the transverse components of the diffracted field are known, the LC $v_z$ behind the screen can be found via equations (2). Performing necessary differentiations in (7), we conclude that equation (7) generates a quite similar relation for the complex amplitude derivatives. Consequently, the LC obeys the same Fresnel-Kirchhoff integral (7); for its importance and suitability of further references, we present it in the explicit form:

$$v_z(x, y, z > 0) = \frac{k}{2\pi i z}\int_a^\infty dy_a \int_{-\infty}^\infty dx_a\, v_z(x_a, y_a, 0)\exp\left\{\frac{ik}{2z}\left[(x-x_a)^2 + (y-y_a)^2\right]\right\}. \tag{8}$$

This relation means that the LC of the paraxial beam field also diffracts independently and its behavior can be studied separately from the transverse ones.

## 3. Diffracted beam trajectory characterization

Thus, we have the sufficient apparatus for calculation of the whole diffracted beam field. In view of the polarization-dependent phenomena, the vortex nature of the LC (6) attracts a special attention. Since it diffracts according to the same rule (8) as the usual transverse one (7), the LC experiences the same vortex-dependent transformations of the diffracted beam profile that were described in many works dealing with the scalar OV beam diffraction (see, e.g., [31–36]). However, since the LC amplitude is normally small with respect to the transverse field, these polarization-dependent features could hardly be discernible in the overall beam intensity profile (unless one employs a specially designed detector selectively sensitive to the longitudinal field, which is a rather exceptional option and requires special conditions [45]). Usually, the beam profile is determined by the total electromagnetic energy distribution [6–12,15–18] and in this case we cannot expect to study all details of the diffracted beam spatial pattern but rather resort to an integral characteristic of its trajectory. As it was accepted in many previous works (see, e.g., [6–12,15,16]), we characterize the beam position by the CG of the energy distribution over the beam cross section

$$\mathbf{r}_c = \frac{\int \mathbf{r} w \, dx \, dy}{\int w \, dx \, dy}$$

where $\mathbf{r} = \begin{pmatrix} x \\ y \end{pmatrix}$ is the transverse radius-vector; from now on, absence of the integration limits means that integration is performed over the whole cross section of the beam. In the free space, the energy density $w$ of the beam electromagnetic field (1), (2) can be represented in the form

$$w = \frac{1}{16\pi}\left(|\mathbf{E}|^2 + |\mathbf{H}|^2\right) = w_\perp + w_z \qquad (9)$$

where the first summand that can be represented in the form

$$w_\perp = \frac{1}{8\pi}\left(|u_x|^2 + |u_y|^2\right) = \frac{1}{4\pi}|u_x|^2 \qquad (10)$$

owes to the transverse components of the field (1), and the second describes the longitudinal field contribution

$$w_z = \frac{1}{16\pi}\left(|v_z|^2 + |v_{Hz}|^2\right) = \frac{1}{8\pi}|v_z|^2 \sim \gamma^2 w_\perp . \qquad (11)$$

Our aim is to inspect the modification of the diffracted beam CG position caused by the change of the incident beam polarization. In agreement with (9), we can separate the contributions of $\mathbf{r}_c$ owing to the longitudinal and transverse field components:

$$\mathbf{r}_c = \mathbf{r}_{c\perp} + \mathbf{r}_{cz} \qquad (12)$$

where

$$\mathbf{r}_{c\perp} = \frac{\int \mathbf{r} w_\perp \, dx \, dy}{\int w \, dx \, dy} \approx \frac{\int \mathbf{r} w_\perp \, dx \, dy}{\int w_\perp \, dx \, dy} = \frac{1}{I}\int \mathbf{r} |u_x|^2 \, dx \, dy \qquad (13)$$

and

$$\mathbf{r}_{cz} = \frac{\int \mathbf{r} w_z \, dx \, dy}{\int w \, dx \, dy} \approx \frac{\int \mathbf{r} w_z \, dx \, dy}{\int w_\perp \, dx \, dy} = \frac{1}{2I}\int \mathbf{r} |v_z|^2 \, dx \, dy . \qquad (14)$$

In equations (13) and (14) $I$ denotes the integral

$$I = \int |u_x|^2 \, dx \, dy ; \qquad (15)$$

the second equalities in (13) and (14) are possible due to the small relative value of $w_z$, see (11).

The quantities (13) and (14) are the main subjects of our further consideration. With knowledge of the evolution of $u_x$ and $v_z$ in the diffracted beam, given by (7) and (8), one can calculate the CG position in arbitrary cross section behind the screen. However, the direct calculations are difficult because the edge-diffracted beam amplitude slowly falls down at $y \to -\infty$, whence the integrals in (13) – (14) diverge, and special limit procedures are necessary to get meaningful results [34]. To avoid these complicated procedures, we employ the fact that in the free space, the CG of the diffracted beam evolves along a rectilinear trajectory [46], i.e., for both summands of (12),

$$\mathbf{r}_{cz}(z) = \mathbf{r}_{cz}(0) + \mathbf{p}_{cz}z, \tag{16}$$

$$\mathbf{r}_{c\perp}(z) = \mathbf{r}_{c\perp}(0) + \mathbf{p}_{c\perp}z \tag{17}$$

where $\mathbf{p}_{cz}$ and $\mathbf{p}_{c\perp}$ are the vectors of angular deviation of the CG [4,46]. To find them we consider the transformation of the paraxial beam complex amplitude on a small path between $z = 0$ and $z = \Delta z$; since the law of $v_z$ evolution (8) is identical to that of the transverse field complex amplitude (7), they can be derived quite similarly. For example, equation (13) of [11] permits us to write

$$u_x(x, y, \Delta z) = u_x(x, y, 0) + \frac{i\Delta z}{2k}\nabla^2 u_x(x, y, 0). \tag{18}$$

This being substituted into (13), after the integration by parts and elementary transformations supposing the complex amplitude properly decays when $|x|, |y| \to \infty$, we arrive at expression (17) with $z \to \Delta z$ in which

$$\mathbf{p}_{c\perp} = \frac{1}{kI}\mathrm{Im}\int u_x^* \nabla u_x \, dx \, dy. \tag{19}$$

Similar operations with $v_z$ and equation (14) give

$$\mathbf{p}_{cz} = \frac{1}{2kI}\mathrm{Im}\int v_z^* \nabla v_z \, dx \, dy. \tag{20}$$

Here the integrations are performed across the screen plane where the difference between the frames $(x_a, y_a)$ and $(x, y)$ (see figure 1) disappears and is no longer reflected in notation. Herewith, the integration domain only includes the 'open' part of the plane (in the 'obscured' part the functions $u_x = v_z = 0$) where the integrand functions behave regularly and contain no improper slowly-decreasing 'tails' associated with the sharp-edge diffraction. Note that equation (19) corresponds to the known expression for the 'tilt' of the CG trajectory [4,11,17] while (20) differs by the additional divider 2 (see (14)).

## 4. Results for the Gaussian beam diffraction

Now we are in a position to make more definite estimates for a concrete experimental situation. Let the incident beam be Gaussian and the screen plane $z = 0$ coincides with its waist plane:

$$u_x(x, y, 0) = u_{00}(x, y) \equiv \exp\left(-\frac{x^2 + y^2}{2b^2}\right) \tag{21}$$

where $b$ is the waist radius measured at the e$^{-1}$ intensity level (the constant amplitude scale factor is omitted as it does not affect the final results), and the subscript '00' refers to the fundamental Laguerre-Gaussian mode [44]. For this beam, the small parameter (3) of the paraxial approximation – the divergence angle $\gamma$ – equals to [4,11,42]

$$\gamma = (kb)^{-1}. \tag{22}$$

Then, according to (6),

$$v_z(x, y, 0) = -i\gamma\frac{x + i\sigma y}{b}\exp\left(-\frac{x^2 + y^2}{2b^2}\right) = -i\gamma\left(\frac{r}{b}\right)e^{i\sigma\phi}u_{00}. \tag{23}$$

Hence, the diffracted field complex amplitudes (7) and (8) can be directly obtained [31,34]

$$u_x(x, y, z > 0) = \frac{1}{1+i\zeta} \exp\left[-\frac{x^2 + y^2}{2b^2(1+i\zeta)}\right] \text{erfc}(\tau),$$

$$v_z(x, y, z > 0) = -i\frac{\gamma}{(1+i\zeta)^2} \exp\left[-\frac{x^2 + y^2}{2b^2(1+i\zeta)}\right]$$

$$\times \left[\frac{x + i\sigma y}{b} \text{erfc}(\tau) + i\sigma \sqrt{\frac{i\zeta(1+i\zeta)}{2\pi}} \exp(-\tau^2)\right] \quad (24)$$

where

$$\text{erfc}(t) = \frac{2}{\sqrt{\pi}} \int_t^\infty \exp(-s^2) ds$$

is the complementary error function [47],

$$\tau = \sqrt{\frac{1+i\zeta}{2i\zeta}} \left(a - \frac{y}{1+i\zeta}\right)$$

and $\zeta = z/kb^2$ is the diffraction beam propagation distance in units of the Rayleigh length [44]. However, in view of the previous section's remark, immediate substitution of (24) into the integral relations (13) and (14) for any current diffraction beam cross section $z > 0$ results in the integrals' divergence. Therefore, characteristics of the CG position (12) can be more suitably derived from relations (16) and (17) that involves (13), (14) and (19), (20) with substitution of the initial ($z = 0$) distributions (21) and (23). Then, the corresponding integrals are to be calculated over the half-space $(-\infty < x < \infty, a < y < \infty)$, as in equations (7) and (8), and can be readily found.

First, we present the transverse field contributions (13) that form the 'background' for small corrections associated with the longitudinal field. The initial shift of the CG position

$$\mathbf{r}_{c\perp}(0) = bG\left(\frac{a}{b}\right) \cdot \begin{pmatrix} 0 \\ 1 \end{pmatrix} \quad (25)$$

is expectably polarization-independent and reflects the effect of partial screening of the beam cross section. Here

$$G(t) = \frac{\exp(-t^2)}{\sqrt{\pi} \, \text{erfc}(t)} \quad (26)$$

is the function whose behavior is illustrated by figure 2. While the beam screening is small (the screen edge is situated at the far periphery of the beam intensity profile ($a \ll -b$), $G(a/b)$ is close to zero, which corresponds to the weak perturbation of the beam. With growing $a$, the CG displaces into the upper half-space, and when the beam is almost completely covered by the screen, $a \gg b$, $G(a/b)$ asymptotically tends to $a/b$, which means that the CG vertical coordinate practically coincides with the screen edge position $a$. The horizontal $x$-component of (25) is always zero due to the incident beam symmetry.

The transverse-field contribution to the trajectory tilt can be determined similarly from (19); we do not need to calculate this quantity since for $u_x$ (21) its $x$-component obviously vanish due to the mirror symmetry with respect to axis $y$, whereas the $y$-component appears to be zero because (21) is a real function; consequently, $\mathbf{p}_{c\perp} = 0$.

The contribution of the longitudinal field to the initial CG shift that follows from (14) and (23),

$$\mathbf{r}_{cz}(0) = b\gamma^2 G\left(\frac{a}{b}\right) \cdot \begin{pmatrix} 0 \\ 1 \end{pmatrix}, \qquad (27)$$

shows a remarkable similarity with the main contribution (25) and, of course, preserves the same symmetry properties. The only difference is the multiplier $\gamma^2$ reflecting the relative intensity of the LC (see (11)); in usual conditions $\gamma \ll 1$ the quantity (27) supplies a very small addition to the background (25). And, at last, the most interesting result is that following from equation (20),

$$\mathbf{p}_{cz} = -\frac{\sigma}{2}\gamma^3 G\left(\frac{a}{b}\right) \cdot \begin{pmatrix} 1 \\ 0 \end{pmatrix}. \qquad (28)$$

This expression (especially, its non-zero $x$-component) describes the main outcome of the paper. It confirms that, indeed, the diffracted beam evolution depends on the incident beam polarization, and the polarization handedness manifests itself in the tiny tilt of the diffracted beam trajectory in the direction parallel to the screen edge. This can be considered as a specific manifestation of the SHE of light in the edge-diffraction process.

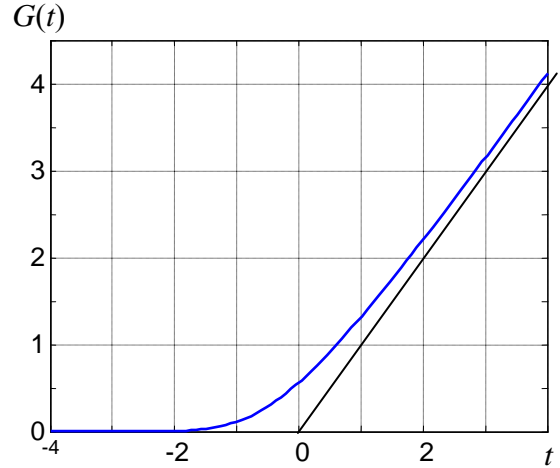

**Figure 2**. (Blue curve) Graph of the function $G(t)$ (26); (black line) asymptote $G = t$.

## 5. Interpretation and discussion

To elucidate the nature of the diffracted beam behavior revealed in the above paragraphs, we consider the field of polarization ellipses in the incident beam cross section; herewith, in contrast to the usual approach, we take the longitudinal field component into account. According to (1), (4) and (21), (23), the total electric field in the incident beam cross section is described by the complex vector

$$\mathbf{E} = u_{00}\left(\mathbf{e}_x + i\sigma\mathbf{e}_y - i\gamma\frac{r}{b}e^{i\sigma\phi}\mathbf{e}_z\right). \qquad (29)$$

It is seen that the longitudinal contribution makes the field polarization slightly elliptical, with different orthogonal semiaxes. Also, the polarization ellipse no longer belongs to the transverse cross section; the normal to its plane is determined by the vector [48]

$$\mathbf{N} = \frac{1}{2}\operatorname{Im}(\mathbf{E}^* \times \mathbf{E}) = \operatorname{Re}\mathbf{E} \times \operatorname{Im}\mathbf{E}$$

(note that $\mathbf{N}$ coincides with the electromagnetic spin density definition [4,39,40]). According to (21), (23) and (29) one obtains

$$\mathbf{N} = u_{00}^2\left[\sigma\mathbf{e}_z + \gamma\frac{r}{b}(\mathbf{e}_y\cos\phi - \mathbf{e}_x\sin\phi)\right] = u_{00}^2\left(\sigma\mathbf{e}_z + \gamma\frac{r}{b}\mathbf{e}_\phi\right) \qquad (30)$$

where $\mathbf{e}_\phi$ is the unit vector of the polar azimuth. This result shows that the polarization ellipses are slightly inclined in the azimuthal direction, as is shown in figure 3, and, since the rotating dipole irradiates most efficiently along the normal to the plane of rotation, this invokes a certain azimuthal 'drift' of the radiation.

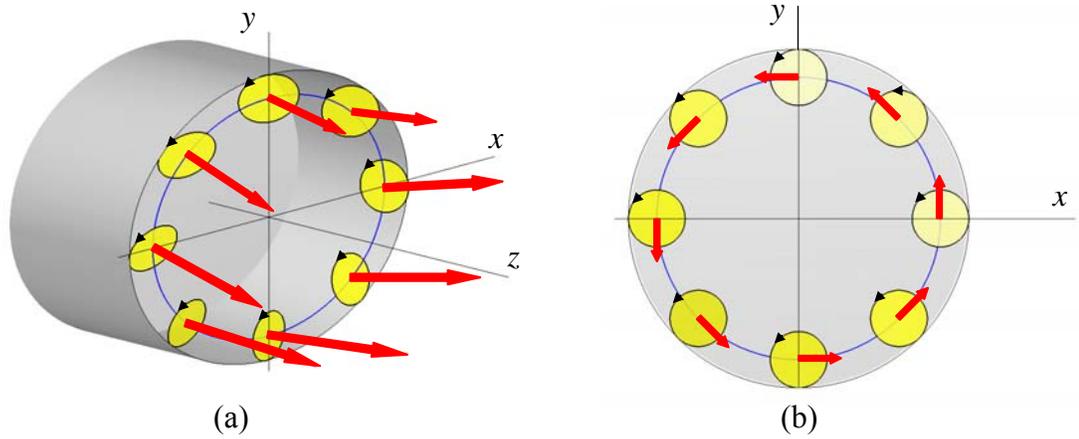

**Figure 3**. Schematic cross-section pattern of a right-polarized ($\sigma = 1$) circular beam with polarization ellipses (yellow) and normals to their planes (red arrows); only ellipses distributed along a single circumference (blue) are shown. Small arrows show the polarization handedness, image (b) represents the frontal view against axis $z$ where the azimuthal inclinations of the normals is clearly seen. Images are calculated via equations (29) and (30), the LC contribution is conventionally enhanced to better visibility of the tilted ellipses.

This scheme can serve an additional explanation of the physical background of the spin energy current [38,39] in a circularly polarized beam. But now more important is that it reveals violation of the symmetry in the radiation distribution with respect to the $y$-axis due to which, when a part of the beam cross section is stopped, the rest propagates with deviation from the 'nominal' $z$-axis direction. In the case of figure 3 (cf. also figure 1) it is the lower part of the beam that is covered, and the azimuthal 'drift' of the upper-part radiation is just directed to negative $x$ and reverses with the sign reversal of $\sigma$, in full agreement with (28).

    The above reasoning provides a realistic qualitative treatment; in this context, the initial motivation of this research based on the analogies with OV beams' diffraction (see Introduction) looks more formal. However, it has permitted us to derive the quantitative description of the discussed phenomenon and reveals its another important aspect illustrated in figure 4. It presents

the far-field intensity distributions of separate field components in the far-field angular coordinates $\alpha_{x,y} = (x, y)/z$ ($z \to \infty$) calculated via the explicit relations (24). The images show that the vortex character of the LC (23) really manifests itself in the diffracted beam propagation, and in quite expectable form (cf., for example, the far-field patterns of the diffracted scalar Laguerre-Gaussian beams [35,36]). While the transverse component of the diffracted field is distributed symmetrically (figure 4(c)), the longitudinal one exhibits articulate spin-dependent lateral shifts, which can be distinctly interpreted as a sort of SHE. However, the possible shifts are small (several $\gamma$ at best). Moreover, because of the extremely small relative weight of the LC magnitude (graphical data for the red and green curves in figure 4(d) should be multiplied by $\gamma^2 \ll 1$), their influence on the total beam energy distribution becomes very weak on the 'huge' symmetric background of the transverse-field contribution. For this reason, we could not present the total diffracted beam deformation in any sensible form and resort to the separate contributions in figures 4(a)–(c).

The relation of the angular deviation of the LC intensity with its vortex nature can also be seen from simple geometric considerations. The LC (6), (23) of the incident beam possesses a helical wavefront, and at a distance $a$ from the axis its azimuthal tilt is [37]

$$\theta = \frac{2\pi/k}{2\pi a} = \frac{1}{ka} = \gamma \frac{b}{a}. \tag{31}$$

When the screen covers the central area of the beam cross section and only the peripheral part of the beam energy still propagates (cf. figure 1), the direction of its propagation is determined by this wavefront tilt, i.e. the trajectory of the LC wave packet deviates in the $x$ direction by the angle (31). Adding the multipliers responsible for the sign of the wavefront inclination, $-\sigma$, and for the relative 'weight' of the LC intensity with respect to the 'background' transverse-field intensity at the beam periphery $r \approx a$, $(\gamma a/b)^2$, we arrive at qualitatively the same result as (28).

In fact, the main difference of the circularly-polarized beam diffraction from the well-known processes associated with the OV-beam diffraction [31–34] is that now these processes take place only in the small LC, which makes the result hardly perceptible and masked by the much more intensive "usual" diffraction of the non-vortex transverse components. This is emphasized by the multiplier $\gamma^3$ in (28). However, although the discussed effect is extremely weak, it is geometrically isolated from other 'background' influences since it is the only $x$-directed component of the CG shift and the only CG trajectory parameter that depends on the incident beam polarization handedness. This enables to hope that this sort of SHE can be detected in an accurate experiment, in which the polarization filtering of the transverse components as well as the polarization-selective means for detection of the longitudinal field [16,45] may be profitable.

For example, following (16) and (28) we can estimate a possible polarization-dependent shift of the beam CG in the plane $z$ behind the screen as $x_c(z) = |\mathbf{p}_{cz}| z \sim \sigma \gamma^3 z$ (for a coarse approximation, all the multipliers of the order of unity are discarded). Evolution of the 'total' beam size in the $x$-direction behind the screen can be evaluated using the Gaussian beam law [44] $b(z) = b\sqrt{1 + (z/kb^2)^2}$, which in the far field $z \gg kb^2$ reduces to $b(z) \sim z/kb = \gamma z$. Accordingly, the expectable relative CG shift due to the polarization-dependent diffraction approximately amounts to

$$\frac{x_c(z)}{b(z)} \sim \sigma \gamma^2. \tag{32}$$

It is a bit disappointing that the shift constitutes a small part of the total beam size but for reasonable paraxial beams with $\gamma \sim 10^{-2} - 10^{-3}$ it seems to be a measurable effect. The situation may be even more favorable with non-paraxial, e.g., strongly focused beams with $\gamma \geq 0.1$ although in that case the above simple theory is not fully applicable.

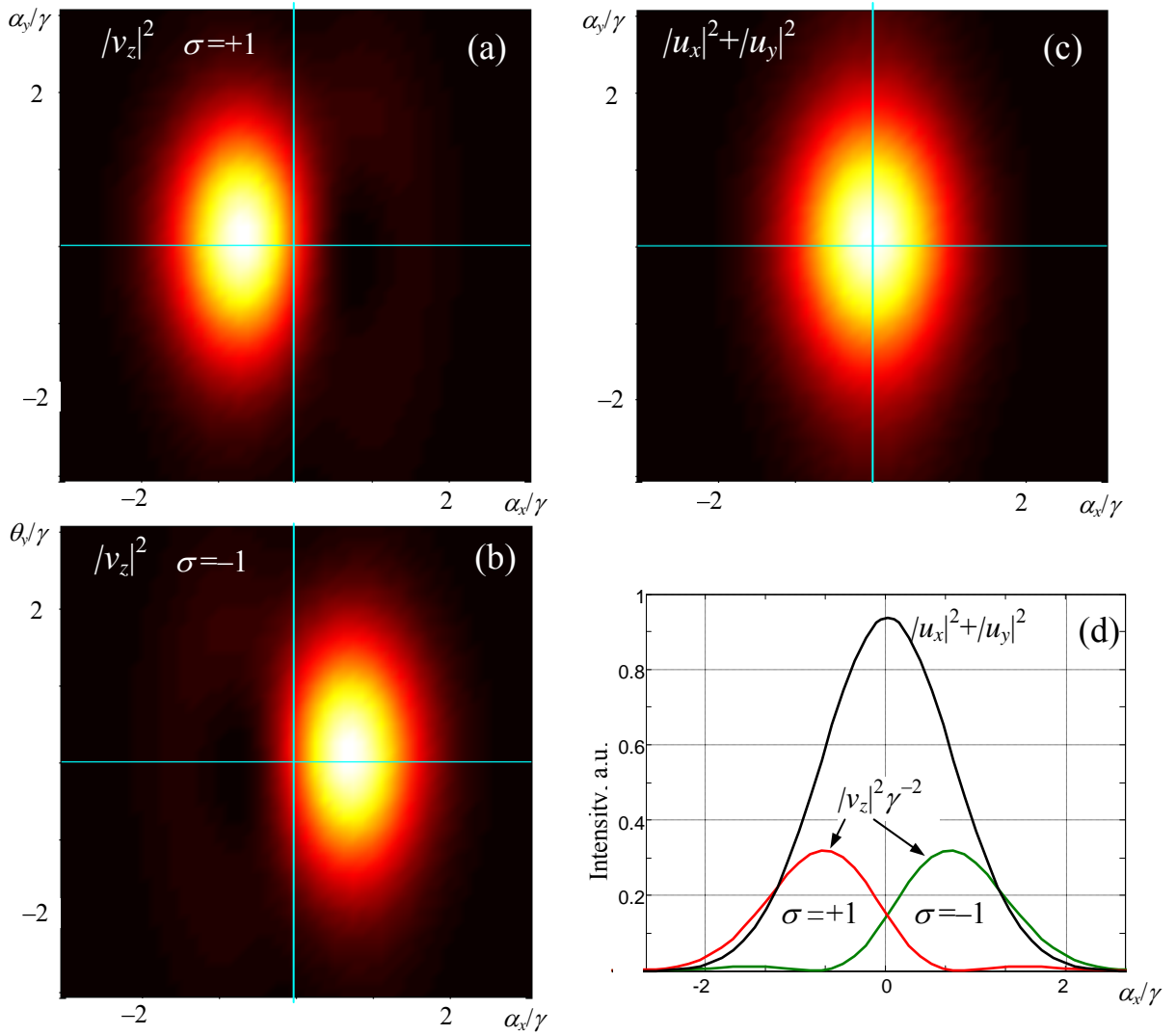

**Figure 4**. Far-field intensity distributions for the diffracted Gaussian beam components (24) for $a = 0$ (the screen edge crosses the incident beam axis, see figure 1 and equations (7), (8)): (a) pseudocolor map of the LC intensity for the incident beam with right circular polarization; (b) the same for the left-polarized incident beam; (c) the same for the transverse field; (d) graphs of the intensity distributions along the $\alpha_x$-axis for: (black) transverse field and (red/green) LC in cases of the right/left incident polarization. In (a) – (c) the transverse coordinates are in units of $\gamma$ (22), in (d) the longitudinal field contributions are magnified by the factor $\gamma^{-2}$.

## 6. Conclusion

In this paper we theoretically consider the edge diffraction of a light beam, based on the paraxial approximation and the Fresnel-Kirchhoff approach, and predict that the spatial pattern of the diffracted beam depends on the incident beam polarization. In particular, if the incident beam is circularly polarized, the trajectory of the diffracted beam centre of gravity (CG) experiences fine angular deviations parallel to the screen edge and reversing the sign with the polarization handedness. This effect is small but important in principle and can be considered as an additional manifestation of the spin-orbit interaction and the spin Hall effect (SHE) of light. It can be interpreted as a revelation of the internal spin energy flow associated with the spin angular momentum in circularly polarized beams and reflects the special role of the incident field longitudinal component that possesses vortex properties.

The peculiar feature of the discussed effect is that it is intimately related to the breakdown of the mirror symmetry with respect to the axis orthogonal to the screen edge which is caused by the circular polarization and internal circulatory flows within the incident beam (see figures 1 and 3). This makes it akin to other symmetry-breaking phenomena originating from the optical-vortex properties and the orbital angular momentum of the incident beam [15–17,32–34,49].

Main features of this effect resemble numerous recent examples of spin-dependent propagation, diffraction, focusing, etc., observed in the specially tailored vector beams with spatially inhomogeneous polarization (see, e.g., [49–51]). However, such tailored beams, regardless of the method of their generation, appear as nothing else than superpositions of opposite circularly polarized modes with spatially inhomogeneous profiles of their amplitude and phase. Actually, the opposite-spin components with different spatial characteristics are always present in the tailored vector beam, though in an implicit form. It is not surprising that after preparation, each mode evolves in its own manner as a usual paraxial beam, and their possible splitting simply reflects the different propagation or diffraction patterns for beams with different initial amplitude and phase profiles. As a result, the observable spin-dependent peculiarities of [49–51] are much larger than those available with 'simple' beams polarized homogeneously [10,12,15–18]. In this view, the effect described in this paper, however weak, looks more fundamental. It characterizes intrinsic properties of any circularly polarized light, independent of the way and scheme of its preparation and emerging only when the first-order terms in the small parameter of paraxial approximation $\gamma$ are involved. The latter circumstance does not make it less essential: the common, ubiquitous and substantial phenomenon of light beam divergence is also a first-order effect in $\gamma$ [42,43].

It should be noted that the SHE version considered in this paper is a purely 'geometrical' effect and is completely determined by the spatial structure of the incident beam and the diffraction obstacle configuration. In fact, some tiny 'material' modifications of the diffracted beam structure may also appear due to the difference in the screen-edge interaction with the $x$- and $y$-polarized field components [44]. Here we did not take into account any 'material' factors that can affect the polarized beam diffraction and, possibly, modify the numerical results. However, these will not break the symmetry discussed in the above paragraphs, so the main features of the polarization-dependent beam shift described in this paper can be refined but not essentially changed. The same arguments are valid in relation to other possible improvements of the results. For example, the initial equations (1) are correct within the first order of the paraxial approximation, i.e. errors of magnitude $\sim\gamma^2$ are possible [42], which seemingly impugns the main result (28). Nevertheless, such errors will not break the right-left symmetry of the beam intensity in figures 1, 3(b), 4(c) and, anyway, the only non-symmetric term $\left(p_{cz}\right)_x$ of (28) remains untouched. Likewise, it can be shown that a possible wavefront spherical curvature (violation of our assumption that the beam waist coincides with the screen plane) can substantially modify the initial CG position (27) as well as introduce the *y*-component of its angular tilt (non-zero transverse field contribution $(p_{c\perp})_y$ and the second element $(p_{cz})_y$ in the column vector (28)) but does not affect the first element $(p_{cz})_x$, which is the only important for the discussed effect.

Similarly, the use of the Gaussian beam model (21) puts no principal limitations for the final result (28); one can expect its validity (within a numerical scaling factor of the order of unity) for any axially-symmetric beam provided that the quantity *b* is a properly defined characteristic of the beam transverse size.

In conclusion we note that the case of the incident beam linear polarization other than horizontal or vertical in the coordinates of figure 1 ($u_y = mu_x$ with real $m \neq 0$) remains beyond the scope of the present work. However, it can be easily analyzed by the same methods. Thinking qualitatively, even just now one can expect that the arbitrary linear polarization will not induce any angular deviation of the diffracted beam CG (the integral in (20) will be a real quantity) but can modify both components of the initial CG shift immediately behind the screen $\mathbf{r}_{cz}(0)$ (16). The expected corrections will be of the same order of magnitude as the studied spin-dependent effect and, probably, hardly measurable because of the strong masking influence of the transverse-field contribution.

**Acknowledgement**

The author is grateful to Petro Maksimyak from Chernivtsi National University, Ukraine, for stimulating ideas.

**Appendix. Orbital Hall effect in diffraction of scalar vortex beams**

The diffraction of scalar OV beams was among the main motives inspiring this research (see section 1). Now we are about to perform the 'feedback': It looks attractive to apply some ideas of the SHE and associated instruments to the scalar diffraction problems. The OV beam diffraction was investigated in much detail [30–36] but except the very specialized work [37] we do not know any attempt to characterize the diffracted beam by a single integral parameter, in contrast to the SHE-oriented researches where the CG is widely used. Here we propose to interpret the known OV-diffraction phenomena in the SHE spirit, i.e. by considering the diffracted beam trajectory in its relation to the incident beam topological charge and OAM. If, as usual, the trajectory is characterized by the CG position $\mathbf{r}_c(z)$, most of the above theory can be directly applied to this case.

For scalar beams, the longitudinal field component is negligible and the transverse polarization is uniform and linear so that the beam can be fully characterized by either of the transverse components (see (1)); we denote its complex amplitude simply by $u(x, y)$, omitting the inessential coordinate subscript. The CG is then fully determined by equations (13) and (15), and its trajectory is described by (17) and (19); all integrals again are calculated over the 'open' part of the beam cross section $-\infty < x < \infty$, $a < y < \infty$.

As an illustration, we apply this scheme to the typical OV beam – circular Laguerre-Gaussian beam with zero radial index and topological charge *l* where

$$u(x,y) = (x \pm iy)^{|l|} \exp\left(-\frac{x^2+y^2}{2b^2}\right) \tag{A1}$$

(for simplicity we again suppose that the screen plane coincides with the beam waist). Substitution of (A1) into equations (15) and (19) yields (omitting the subscript "⊥")

$$\mathbf{p}_c = \begin{pmatrix} p_c \\ 0 \end{pmatrix}, \quad p_c = -\frac{l}{kI} \int_{-\infty}^{\infty} dx \exp\left(-\frac{x^2}{b^2}\right) \int_a^{\infty} y(x^2+y^2)^{|l|-1} \exp\left(-\frac{y^2}{b^2}\right) dy, \tag{A2}$$

$$I = \int_{-\infty}^{\infty} dx \exp\left(-\frac{x^2}{b^2}\right) \int_a^{\infty} (x^2+y^2)^{|l|} \exp\left(-\frac{y^2}{b^2}\right) dy. \tag{A3}$$

Expectedly, the angular deviation of the diffracted beam trajectory is parallel to the screen edge and proportional to the incident OV topological charge. In case $l = \pm 1$, a simple analytical representation of (A2) is available:

$$p_c = \mp \gamma \frac{b}{a}\left[1 + \sqrt{\pi}\frac{b}{a}\mathrm{erfc}\left(\frac{a}{b}\right)\exp\left(\frac{a^2}{b^2}\right)\right]^{-1} \tag{A4}$$

where $\gamma$ is the self-divergence angle (22) of the Gaussian envelope of the beam (A1). For example, if $a = 0$ (the screen edge crosses the beam axis), (A4) gives $p_c = \mp \gamma/\sqrt{\pi}$. This result reflects the magnitude of the discussed effect: generally $p_{cx}$ is of the order of $\gamma$. Asymptotically, for $a \gg b$, expression (A4) reduces to $p_c \simeq \mp \gamma(b/a)$ which remarkably corresponds to the wavefront inclination (31). Like (28), the angular deviation (A4) vanishes at large negative $a$ but, in contrast to (28), its magnitude does not grow monotonically with $a$ and possesses a maximum $|p_c| \approx 0.63\gamma$ at $a \approx 0.43b$.

For arbitrary topological charge $l$, explicit analytical representation of the result similar to (A2), (A3) is cumbersome but it can be simplified in a few special situations. First, at $a \to -\infty$, obviously $p_c \to 0$ due to the symmetry of the internal integral in (A2); second, in the asymptotic limit $a \gg b$ expressions (A2), (A3) can be evaluated approximately, which lead to the simple and physically transparent expression

$$p_c \simeq -\frac{l}{ka} = -l\gamma\left(\frac{b}{a}\right). \tag{A5}$$

Like (31), this result associates the trajectory direction with the local azimuthal tilt of the wavefront but reflects its growth with the topological charge of the OV beam. And the third solvable situation occurs at $a = 0$ when the double integrals (A2) and (A3) can be exactly calculated in polar coordinates, and the result reads

$$p_c = -\frac{2l}{\pi k b}\frac{\Gamma\left(|l|+\frac{1}{2}\right)}{\Gamma(|l|+1)}$$

where $\Gamma$ is the gamma-function symbol [47].

This appendix gives a little addition to the accumulated knowledge of the OV beams' diffraction but it exposes some new aspects which are interesting for comparison with the circularly-polarized beam diffraction described in the main text and, as far as we can judge, were never properly addressed in the current literature.

**References**

[1] Bliokh K Y, Niv A, Kleiner V and Hasman E 2008 Geometrodynamics of spinning light *Nature Photon.* **2** 748–53
[2] Bliokh K Y, Ostrovskaya E A, Alonso M A, Rodríguez-Herrera O G, Lara D and Dainty C 2011 Spin-to-orbital angular momentum conversion in focusing, scattering, and imaging systems *Opt. Express* **19** 26132–49
[3] Bliokh K Y, Rodríguez-Fortuño F J, Nori F and Zayats A V 2015 Spin-orbit interactions of light *Nature Photon.* **9** 796–808
[4] Bekshaev A Bliokh K and Soskin M 2011 Internal flows and energy circulation in light beams *J. Opt.* **13** 053001
[5] Marrucci L, Karimi E, Slussarenko S, Piccirillo B, Santamato E, Nagali E and Sciarrino F 2011 Spin-to-orbital conversion of the angular momentum of light and its classical and quantum applications *J. Opt.* **13** 064001
[6] Liberman V S and Zel'dovich B Y 1992 Spin-orbit interaction of a photon in an inhomogeneous medium *Phys. Rev.* A **46** 5199–207
[7] Fedoseyev V G 1991 Lateral displacement of the light beam at reflection and refraction *Opt. Spektrosk.* **71** 829-34 [*Opt. Spectr.* (USSR) **71** 483]; *Opt. Spektrosk.* **71** 992–997 [*Opt. Spectr.* (USSR) **71** 570]
[8] Onoda M, Murakami S and Nagaosa N 2004 Hall effect of light *Phys. Rev. Lett.* **93** 083901
[9] Bliokh K Y and Bliokh Y P 2007 Polarization, transverse shifts, and angular momentum conservation laws in partial reflection and refraction of an electromagnetic wave packet *Phys. Rev.* E **75** 066609
[10] Bliokh K Y and Aiello A 2013 Goos–Hänchen and Imbert–Fedorov beam shifts: an overview *J. Opt.* **15** 014001
[11] Bekshaev A Ya 2012 Polarization-dependent transformation of a paraxial beam upon reflection and refraction: A real-space approach *Phys. Rev.* A **85** 023842
[12] Hosten O and Kwiat P 2008 Observation of the spin Hall effect of light via weak measurements *Science* **319** 787–90
[13] Bliokh K Y 2009 Geometrodynamics of polarized light: Berry phase and spin Hall effect in a gradient-index medium *J. Opt. A: Pure Appl. Opt.* **11** 094009
[14] Marrucci L, Manzo C and Paparo D 2006 Optical spin-to-orbital angular momentum conversion in inhomogeneous anisotropic media *Phys. Rev. Lett.* **96** 163905
[15] Baranova N B, Savchenko A Y and Zel'dovich B Y 1994 Transverse shift of a focal spot due to switching of the sign of circular polarization *JETP Lett.* **59** 232–34
[16] Zel'dovich B Y, Kundikova N D and Rogacheva L F 1994 Observed transverse shift of a focal spot upon a change in the sign of circular polarization *JETP Lett.* **59** 766–69
[17] Bekshaev A 2011 Improved theory for the polarization-dependent transverse shift of a paraxial light beam in free space *Ukr. J. Phys. Opt.* **12** 10–18
[18] Aiello A, Lindlein N, Marquardt C and Leuchs G 2009 Transverse angular momentum and geometric spin Hall effect of light *Phys. Rev. Lett.* **103** 100401
[19] Zhao Y, Edgar J S, Jeffries G D M, McGloin D and Chiu D T 2007 Spin-to-orbital angular momentum conversion in a strongly focused optical beam *Phys. Rev. Lett.* **99** 073901
[20] Bliokh K Y 2006 Geometrical optics of beams with vortices: Berry phase and orbital angular momentum Hall effect *Phys. Rev. Lett.* **97** 043901
[21] Fadeyeva T A, Rubass A F and Volyar A V 2009 Transverse shift of a high-order paraxial vortex-beam induced by a homogeneous anisotropic medium *Phys. Rev. A* **79** 053815

[22] Fedoseyev V G 2001 Spin-independent transverse shift of the centre of gravity of a reflected and of a refracted light beam *Opt. Commun.* **193** 9–18
[23] Fedoseyev V G 2008 Transformation of the orbital angular momentum at the reflection and transmission of a light beam on a plane interface *J. Phys. A: Math. Theor.* **41** 505202
[24] Okuda H and Sasada H 2008 Significant deformations and propagation variations of Laguerre-Gaussian beams reflected and transmitted at a dielectric interface *J. Opt. Soc. Am. A* **25** 881–90
[25] Okuda H and Sasada H 2006 Huge transverse deformation in nonspecular reflection of a light beam possessing orbital angular momentum near critical incidence *Opt. Express* **14** 8393–8402
[26] Bekshaev A Ya and Popov A Yu 2002 Method of light beam orbital angular momentum evaluation by means of space-angle intensity moments *Ukr. J. Phys. Opt.* **3** 249–57
[27] Dasgupta R and Gupta P K 2006 Experimental observation of spin-independent transverse shift of the centre of gravity of a reflected Laguerre-Gaussian light beam *Opt. Commun.* **257** 91–96
[28] Bekshaev A Ya 2009 Oblique section of a paraxial light beam: criteria for azimuthal energy flow and orbital angular momentum *J. Opt. A: Pure Appl. Opt.* **11** 094003
[29] Bekshaev A 2011 Role of azimuthal energy flows in the geometric spin Hall effect of light (arXiv:1106.0982)
[30] Marienko I G, Vasnetsov M V and Soskin M S 1999 Diffraction of optical vortices *Proc. SPIE* **3904** 27–34
[31] Masajada J 2000 Half-plane diffraction in the case of Gaussian beams containing an optical vortex *Opt. Commun.* **175** 289–94
[32] Arlt J Handedness and azimuthal energy flow of optical vortex beams 2003 *J. Mod. Opt.* **50** 1573–80
[33] Cui H X, Wang X L, Gu B, Li Y N, Chen J and Wang H T 2012 Angular diffraction of an optical vortex induced by the Gouy phase *J. Opt.* **14** 055707
[34] Bekshaev A Ya, Mohammed K A and Kurka I A 2014 Transverse energy circulation and the edge diffraction of an optical-vortex beam *Appl. Opt.* **53** B27–37
[35] Bekshaev A and Mohammed K A 2013 Transverse energy redistribution upon edge diffraction of a paraxial laser beam with optical vortex *Proc. SPIE* **9066** 906602
[36] Bekshaev A Ya and Mohammed K A 2015 Spatial profile and singularities of the edge-diffracted beam with a multicharged optical vortex *Opt. Commun.* **341** 284–94
[37] Bekshaev A Ya, Kurka I A, Mohammed K A and Slobodeniuk I I 2015 Wide-slit diffraction and wavefront diagnostics of optical-vortex beams *Ukr. J. Phys. Opt.* **16** 17–23
[38] Bekshaev A Ya and Soskin M S 2007 Transverse energy flows in vectorial fields of paraxial beams with singularities *Opt. Commun.* **271** 332–48
[39] Berry M V 2009 Optical currents *J. Opt. A: Pure and Applied Optics* **11** 094001
[40] Bliokh K Y, Bekshaev A Y and Nori F 2013 Dual electromagnetism: helicity, spin, momentum and angular momentum *New J. Phys.* **15** 033026
[41] Bliokh K Y and Nori F 2015 Transverse and longitudinal angular momenta of light. *Phys. Reports* **592** 1–38
[42] Lax M, Louisell W H and McKnight W B 1975 From Maxwell to paraxial wave optics *Phys. Rev.* A **11** 1365–70
[43] Bekshaev A Ya and Grimblatov V M 1989 Violation of transversality and transfer of electromagnetic field energy in coherent light beams *Optics and Spectroscopy* **66** 127–8

[44] Solimeno S, Crosignani B and DiPorto P 1986 *Guiding, Diffraction and Confinement of Optical Radiation* (Orlando: Academic Press)
[45] Hayazawa N, Saito Y and Kawata S 2004 Detection and characterization of longitudinal field for tip-enhanced Raman spectroscopy *Appl. Phys. Lett*. **85** 6239–41
[46] Anan'ev Yu A and Bekshaev A Ya 1994 Theory of intensity moments for arbitrary light beams *Opt. Spectr*. **76** 558–68
[47] Abramovitz M and Stegun I 1964 *Handbook of mathematical functions* (National Bureau of standards, Applied mathematics series, 55)
[48] Berry M V and Dennis M R 2001 Polarization singularities in isotropic random vector waves *Proc. R. Soc. Lond. A* **457** 141–55
[49] Ling X, Yi X, Zhou X, Liu Y, Shu W, Luo H and Wen S 2014 Realization of tunable spin-dependent splitting in intrinsic photonic spin Hall effect *Appl. Phys. Lett.* **105** 151101
[50] Zhou J, Zhang W, Liu Y, Ke Y, Liu Y, Luo H and Wen S 2016 Spin-dependent manipulating of vector beams by tailoring polarization *Scientific Reports* **6** 34276
[51] Ling X, Zhou X, Huang K, Liu Y, Qiu C W, Luo H and Wen S 2017 Recent advances in the spin Hall effect of light *Reports on Progress in Physics* **80** 066401